\documentclass[fleqn,10pt]{wlscirep}
\usepackage[utf8]{inputenc}
\usepackage[T1]{fontenc}
\usepackage{amsmath} 
\usepackage{booktabs} 
\usepackage{tabularx} 
\usepackage{float} 
\usepackage{amsmath}
\usepackage{algorithm}
\usepackage{algorithmic}
\title{Laser Scan Path Design for Controlled Microstructure in Additive Manufacturing with Integrated Reduced-Order Phase-Field Modeling and Deep Reinforcement Learning}

\author[1]{Augustine Twumasi}
\author[2]{Prokash Chandra Roy}
\author[3]{Zixun Li}
\author[2]{Soumya Shouvik Bhattacharjee}
\author[3,*]{Zhengtao Gan}
\affil[1]{Computational Science Program, The University of Texas at El Paso, El Paso, TX, 79968}
\affil[2]{Department of Aerospace and Mechanical Engineering, The University of Texas at El Paso, El Paso, TX, 79968}
\affil[3]{School of Manufacturing Systems and Networks, Arizona State University, Mesa, AZ, 85212}

\affil[*]{Corresponding author email: zhengtao.gan@asu.edu}

\keywords{Reduced order modelling (ROM), Deep Reinforcement Learning(DRL), Phase-field Modelling(PFM), Microstructure Evolution, Additive Manufacturing}

\begin{abstract}

Laser powder bed fusion (L-PBF) is a widely recognized additive manufacturing technology celebrated for its ability to produce intricate metal components with exceptional accuracy. However, a notable challenge in L-PBF is the formation of complex microstructures, which significantly affects the quality of the final product. To address this challenge, we propose a physics-guided, machine-learning-aided approach to optimize scan paths to obtain desired microstructure outcomes, such as the development of equiaxed grains to enhance material properties. To gain insights into microstructure evolution, we utilized a phase-field method(PFM) to model the evolution of crystalline grain structure. To reduce computational costs, we trained a surrogate machine learning model, i.e, a 3D U-Net convolutional neural network, using a dataset including single-track phase-field simulations with various laser powers. This enables the machine learning model to accurately predict crystalline grain orientations based on the initial microstructure and thermal history. We conducted an initial investigation of three primary scanning strategies across various hatch spacings within a square domain, demonstrating a speedup by at least two orders of magnitude using the surrogate model. Furthermore, to reduce trial and error in designing laser scan toolpaths, we leverage deep reinforcement learning (DRL) to guide laser movement and generate optimized scan paths that achieve the target microstructure. Results from three distinct cases highlight the effectiveness of the DRL approach. We integrated the surrogate 3D U-Net model into our DRL environment set up to accelerate the reinforcement learning training process. The designed reward function focuses on minimizing both the aspect ratio and the grain volume of the predicted microstructure resulting from the agent’s scan path. The reinforcement learning algorithm was benchmarked against the conventional zigzag approach for both smaller and larger domains, demonstrating the potential of machine learning methods to not only enhance the precision of microstructure control but also significantly improve computational efficiency in L-PBF optimization.

\end{abstract}

\begin{document}
\maketitle
%
%


\section{INTRODUCTION}

Laser powder bed fusion (L-PBF) has emerged as a leading next-generation fabrication technique, renowned for its ability to produce intricate geometric components with remarkable precision. This process involves using a high power laser to melt metal powders layer-by-layer, culminating in fully dense metallic parts \cite{jadhav2021laser, ding2022geometric, qu2021high}. Despite its advantages, L-PBF faces several challenges. Post-build electron backscatter diffraction (EBSD) analysis has revealed irregular polycrystalline microstructures, featuring a combination of equiaxed and elongated columnar grains oriented in various directions \cite{andreau2019texture, leicht2020effect, popovich2017functionally}. The microstructural characteristics of L-PBF parts are significantly influenced by a range of process parameters, including reheating temperature, laser power, scanning speed, powder properties, and scanning strategies \cite{vikram2023effect, gu2020effects, evangelou2023effects}.

The choice of scan pattern plays a crucial role in influencing local thermal conditions and thermal stress, which strongly impact the microstructure formation \cite{boissier2020additive, jhabvala2010effect, ma2007temperature}. The study by \cite{vcapek2022influence} on Alloy 718 demonstrates the significant impact of laser scanning paths on microstructure, using neutron diffraction, EBSD, and synchrotron X-ray diffraction to characterize crystallographic texture, grain morphology, and microstructural heterogeneity. Ekubaru et al. \cite{ekubaru2022effects} emphasized the crucial role of scanning strategies in L-PBF, showing that X-scan strategies in the Sc-Zr-modified Al-Mg alloy led to superior mechanical properties due to the increased volume fraction of ultrafine grains (UFGs) along the melt pool boundaries, enhancing tensile and yield strengths. Zigzag scan patterns are commonly employed due to their stability and simplicity, but they can intensify thermal stress and cause excessive distortion \cite{zhao2020shape, ramos2019new, qiu2013microstructure}. Previous studies by Yang et al. \cite{yang2003fractal} and Catchpole-Smith et al. \cite{catchpole2017fractal} investigated fractal scanning patterns with limited theoretical underpinning and practical implementation. The identification of thermal gradients as a primary cause of residual stress and distortion, which affects microstructure, has led to several toolpath optimization methods based on thermal-mechanical models \cite{yan2018stress}. Boissier et al. \cite{boissier2022time} examined temperature field variations in scan path optimization, but their approach resulted in variable hatch spacing, leading to non-uniform or incomplete melting. 

Chen et al. \cite{chen2021island} proposed an island-based scan pattern generation method using finite element analysis (FEA) with approximated voxels. While effective, this method is computationally intensive, especially with complex cross-sectional geometries, resulting in heterogeneous microstructures influenced by localized heating and cooling rates. Ramani et al. \cite{ramani2022smartscan} introduced the "smartscan" method to enhance thermal uniformity in zigzag and island patterns, improving microstructural consistency by reducing thermal gradients. Chen et al. \cite{chen2020level} developed a continuous laser scanning optimization method to reduce residual stress for specific geometries, showing promise in simulations. Takezawa et al. \cite{takezawa2022simultaneous} optimized hatching orientation and lattice density distribution, validated through simulations and experiments, but still relied on traditional zigzag scan patterns. Qin et al. \cite{qin2023adaptive} proposed adaptive toolpath generation (ATG) algorithms to minimize thermal gradients by selecting the "best" next point within the search regions. However, this approach sometimes resulted in extreme temperature accumulation, potentially leading to undesirable microstructural features such as large columnar grains. 

Computational methods, such as Phase-field method \cite{sahoo2016phase, chu2020phase}, Cellular Automaton \cite{teferra2021optimizing}, and Kinetic Monte Carlo \cite{rodgers2017simulation}, have been used to simulate microstructure evolution but they are computationally expensive due to the high spatial and temporal resolutions required for the accurate modeling. Data-driven machine learning models have been developed as reduced-order models that can significantly accelerate the simulations. Methods such as autocorrelation, principal component analysis (PCA), physics-embedded graph networks \cite{xue2022physics}, and convolutional neural networks (CNNs), specifically the 3D U-Net \cite{ronneberger2015u, choi4502223accelerating}, have shown promise in this context. These models reduce microstructure data to a low-dimensional representation, significantly accelerating physics-based simulations. Recent research has explored deep reinforcement learning (DRL) applications in metal additive manufacturing (AM). Dharmawan et al. \cite{dharmawan2020model} introduced a reinforcement learning-based control strategy to improve surface quality in robotic metal wire arc AM. Their approach treated toolpath points as agents following zigzag scan patterns but did not focus on optimizing thermal distribution or microstructural evolution. Ogoke and Farimani \cite{ogoke2021thermal} explored DRL algorithms as an effective control policy for thermal management in the LPBF process, primarily optimizing parameters like laser power and velocity. Qin et al. \cite{qin2024deep} developed a DRL-based toolpath generation framework to address residual stress but restricted their study to turning angles, leaving broader implications for microstructure evolution less explored.



\begin{figure}[H]
    \centering
    \includegraphics[width=0.95\linewidth]{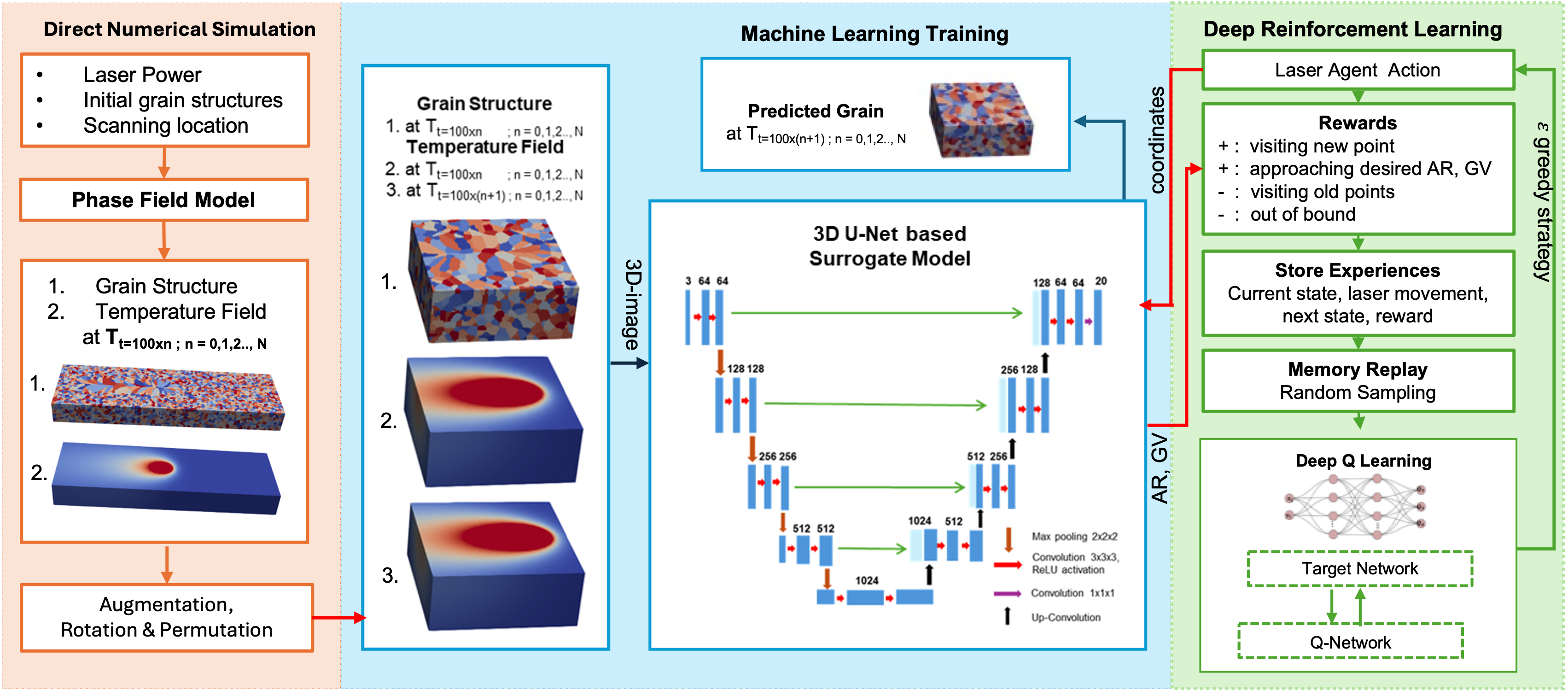}
    \caption{Overview of the Methodology: The figure illustrates our framework, which integrates Direct Numerical Simulation(DNS) for high-fidelity grain morphology prediction using phase-field method, a 3D-UNet (ROM) for learning spatial features and predicting grain evolution, and DRL for optimizing scan paths in the L-PBF process. DNS provides ground truth data for training the 3D-UNet, while DRL utilizes the learned features from the 3D-UNet to adaptively adjust the laser path, thereby enhancing efficiency and grain structure quality.}
    \label{fig-scheme}
\end{figure}

In this paper, our proposed approach builds upon the pipeline introduced in our previous work by \cite{twumasi2024brief}, focusing on optimizing laser scan paths in L-PBF through the application of deep reinforcement learning. This method is further enhanced by the integration of an accelerated machine learning model (3D U-Net), as illustrated in Figure \ref{fig-scheme}. Figure \ref{fig-scheme} includes data sampling through single-track phase-field simulations, training a machine learning model using a 3D-UNet, and subsequently integrating the reduced-order ML model into the DRL framework. The DRL algorithm encompasses uniform sampling, agent movement, environment observation, action selection, movement constraints, reward calculation, and training. The DRL-based algorithm aims to minimize the grain aspect ratio and grain volume by leveraging the computational efficiency of the 3D-UNet model for fast predictions of grain growth. The action spaces are defined with four possible movements: up, down, left, or right, all designed to navigate the high degrees of freedom inherent in the scan pattern design space, with the goal of minimizing the quantities of interest, i.e., grain volume and aspect ratio. We compared DRL-designed tool path with standard zig-zag scan pattern in terms of average aspect ratio and grain volume. Additionally, we extended our analysis to a larger domain simulation, demonstrating the effectiveness and the scalability of the proposed model. 


\section{RESULTS AND DISCUSSION}
\subsection{Machine Learning Prediction of Grain Structure with Heuristic Laser Tool Paths}
\begin{figure}[H]
    \centering
    \includegraphics[width=0.95\linewidth]{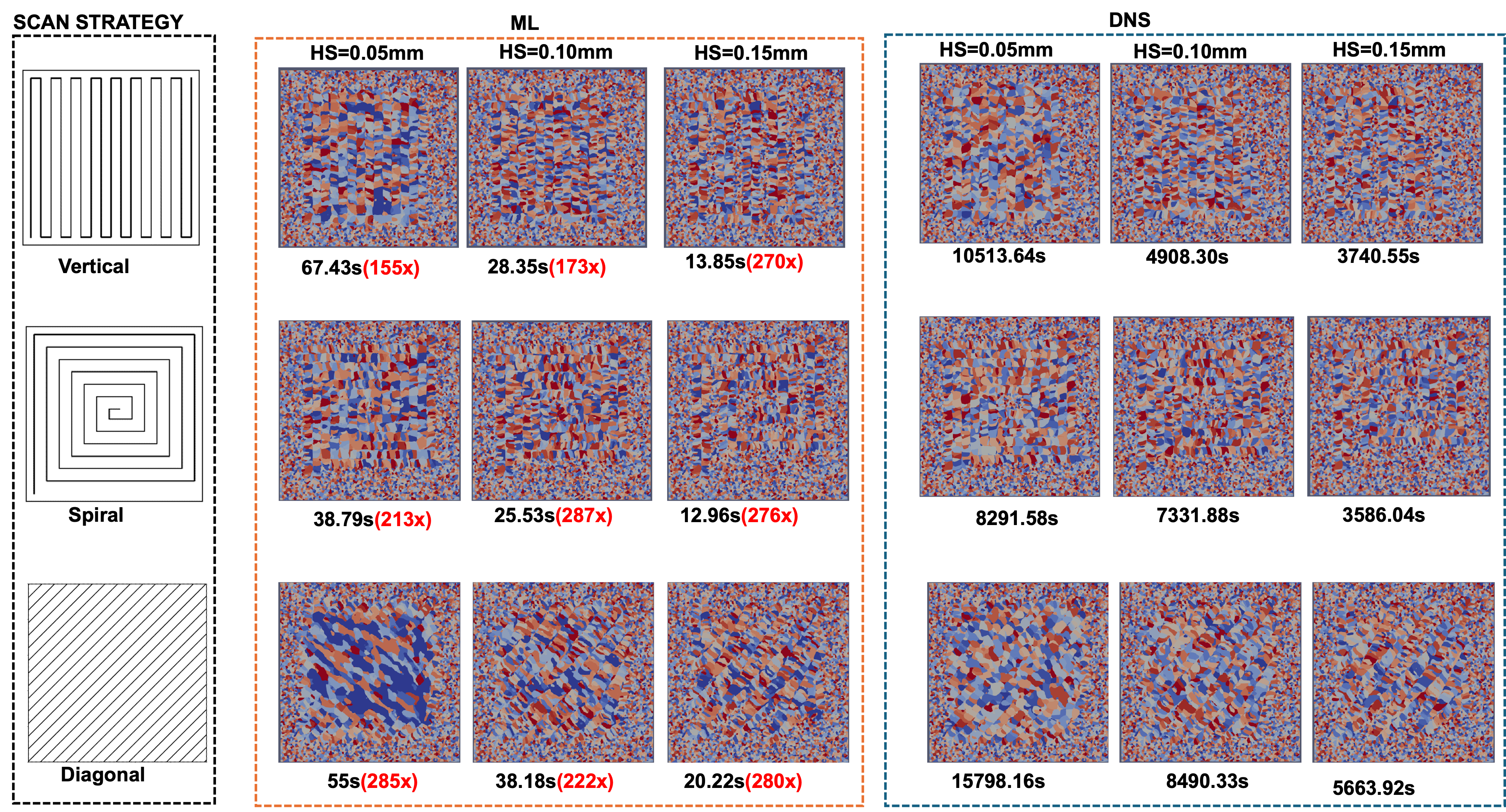}
    \caption{Illustration of three different scan strategies (vertical serpentine, spiral serpentine, Diagonal) for ML and DNS highlighting their respective speed-up gains.  The ML model achieves a computational speed-up of up to three orders of magnitude over DNS }
    \label{fig-scan}
\end{figure}

The trained ML model was evaluated on single, two, and three-track configurations using a new initial grain structure and a heat input of $Q=25W$, not present in the training data. Comparisons with direct numerical simulations (DNS) reveal that the model's predictions qualitatively align with DNS results, particularly on the top surface grain formations. 
We exhibit the forecasting capabilities of our machine learning model under conditions of limited computational resources by evaluating its performance across various scan patterns. Specifically, we examine three distinct patterns, namely the Vertical Serpentine Pattern, characterized by the laser moving in a vertical zigzag fashion across the build area; the clockwise spiral pattern, where the laser spirals in a clockwise direction; and the diagonal  Pattern, which involves diagonal laser scanning each with varied hatch spacings as shown in the left region of Figure \ref{fig-scan}, resulting in a comprehensive analysis of nine scenarios with a domain size of $1\ mm \times 1\ mm \times 0.1\ mm$. 
Our intention is to analyze the model's performance across these diverse patterns in order to establish its potential as a viable alternative to computationally intensive DNS, to speedup the reinforcement learning process.

Figure \ref{fig-scan} depict the predicted grain structure at the final time step for different scan patterns for both the ML and the DNS. The qualitative agreement between the ML model's predictions and those of the DNS is evident across most scan patterns, with minor differences observed. This alignment demonstrates the robustness of the ML model in capturing the intricate details of grain structure formation under various scanning strategies.
However, significant discrepancies between the ML and DNS predictions are apparent in the diagonal part with hatch spacing of $0.05\ mm$. These differences can be attributed to the training data being sampled from single-track scans without overlapping melt pool tracks in a smaller domain. The ML model was trained on this limited dataset and then extrapolated to predict grain structures in a larger domain with multiple overlapping tracks. This extrapolation likely introduced errors, especially in regions with complex melt pool interactions.

As the hatch spacing increases, the prediction accuracy improves, suggesting that the model performs better in situations with less overlapping and more distinct track separations. This observation indicates that while the ML model is effective in simpler scenarios, its accuracy decreases in more complex, high-overlap situations.
Despite these limitations, the model's ability to generalize from a limited training set to broader applications is noteworthy. It provides valuable predictive insights into grain structure development influenced by varying hatch strategies in L-PBF processes.

It is important to mention that the speed-up achieved through ML methods is approximately two to three orders of magnitude greater than that of DNS approaches as shown in red scheme. 
This findings  shows an inverse relationship between hatch spacing and simulation time in  L-PBF processes. Particularly, wider hatch spacings are associated with a substantial decrease in computational time. The reason for this efficiency lies in the fact that the laser covers a larger area per pass, thus reducing the number of passes required. This correlation has been observed in both ML and DNS simulations, and the ML methods have shown a particularly remarkable increase in computational efficiency. which underscores the superior efficiency of ML methods.

Additionally, we performed a quantitative comparison of grain morphology predictions using ML and DNS. The effectiveness of our model in predicting the desired microstructure for L-PBF may not always be fully reflected in standard metrics, such as loss and voxel accuracy. To address this limitation, we employ a morphology-similarity metric that focuses on grain volume and aspect ratios, enhancing decision making in scan path selection. We utilized a breadth-first search (BFS) to identify connected voxels with the same orientation, treating them as single grains. The grain volume was calculated by dividing the total volume of the identified grains by their counts. The aspect ratio, denoted as $\Phi = \frac{2a}{b+c}$, is the ratio of the longer axis 'a' to the sum of the shorter axes ('b' and 'c') of an equivalent ellipsoid. To accurately compare the significantly grown grains, those with a volume less than the initial grain volume of $500\ \mu m^{3}$ were excluded from the aspect ratio analysis. The Root Mean Square Error (RMSE) and Normalized Root Mean Square Error (NRMSE) were used to quantify discrepancies in the grain volume and aspect ratios between the ML model's predictions and the DNS results for various scan paths. 

Table \ref{table:comparison} shows that both the Root Mean Squared Error (RMSE) and Normalized Root Mean Squared Error (NRMSE) decrease as the hatch spacing increases. 



\begin{table}[H]
\caption{Comparative RMSE and NRMSE for Volume and Aspect Ratio Estimations by ML and DNS Across Scan Patterns.}
    \centering
    \begin{tabular}{@{}lcccccc@{}}
        \toprule
        Scan Paths & Hatch Space (mm) & \multicolumn{2}{c}{Volumes} & \multicolumn{2}{c}{Aspect Ratio} \\
        \cmidrule(r){3-4} \cmidrule(l){5-6}
                   & & RMSE & NRMSE & RMSE & NRMSE \\
        \midrule
        Diagonal   & 0.05 & 1334 & 0.0231 & 1162 & 0.0239 \\
                   & 0.10 & 1052 & 0.0171 & 894 & 0.0170 \\
                   & 0.15 & 828 & 0.0129 & 703 & 0.0129 \\
        Vertical   & 0.05 & 700 & 0.0117 & 557 & 0.0110 \\
                   & 0.10 & 667 & 0.0105 & 513 & 0.0095 \\
                   & 0.15 & 612 & 0.0093 & 485 & 0.0087 \\
        Spiral     & 0.05 & 613 & 0.0101 & 472 & 0.0092 \\
                   & 0.10 & 673 & 0.0105 & 506 & 0.0092 \\
                   & 0.15 & 545 & 0.0082 & 424 & 0.0075 \\
        \bottomrule
    \end{tabular}
    \label{table:comparison}
\end{table}

\subsection{Inverse Tool Path Design using Deep Reinforcement Learning}

In the previous section, we discussed the ML-based forward prediction of grain structure using heuristic tool paths. In this section, we explore inverse tool path design aimed at achieving target microstructure features, leveraging the deep reinforcement learning. To optimize toolpaths for achieving the desired grain morphology, the reward function in our DRL method incorporates both aspect ratio (AR) and grain volume (GV). 

AR and GV significantly influence the properties of parts produced using L-PBF. Specifically, a reduction in AR enhances the isotropic mechanical performance of L-PBF Ti-6Al-4V alloys \cite{drl2}, leading to improved industrial applicability.
Additionally, reducing AR increases the compressive strength and relative density of SLM Ti-6Al-4V alloy lattice materials \cite{drl4}. Smaller grains or grains with lower GV provide higher strength but reduced ductility \cite{drl5, drl6}, whereas larger grains enhance ductility. Grain size also affects dynamic mechanical properties such as fracture toughness and grain propagation \cite{drl7}. Thus, controlling grain morphology is crucial in the LPBF process. While desired grain morphology can also be achieved through post-processing heat treatments such as Hot Isostatic Pressing (HIP), annealing, or quenching, our approach aims to directly optimize toolpaths to facilitate better morphology before any thermal treatment. 

In our DRL approach, we explored three distinct reward function cases with \textbf{case 1} and \textbf{case 2} are for a small domain while \textbf{case 3} is for a larger domain:
 \begin{enumerate}
     \item \textbf{Case 1:} The reward function is based solely on aspect ratio, aiming to find an optimized path that minimizes the average aspect ratio after scanning.
     \item \textbf{Case 2:} The reward function focuses on grain volume, with the goal of finding a path that results in a lower average grain volume after scanning.
     \item \textbf{Case 3:} The reward function combines both aspect ratio and grain volume, seeking an optimized path where both metrics are minimized across the entire prescribed region.
 \end{enumerate}
Figure \ref{case1} presents the results for \textbf{Case 1} and \textbf{Case 2}, from the implementation of the proposed DRL algorithm.
\subsubsection{Case 1: Using Aspect Ratio as a Reward Function}


Case 1 of figure \ref{case1} shows the results of  minimizing the average aspect ratio (AR) using our DRL method. The figure \ref{case1}(d) plot illustrates the cumulative rewards over episodes, while figures \ref{case1}(b) depict the evolution of scan paths over time, showing the laser agent’s progressive attempts to optimize the path coverage.
Initially, the cumulative reward, comprising of weighted penalties for average aspect ratio (AR) and grain volume (GV), as well as penalties for collision/remelting, moving out of bounds, and neglecting parts of the simulation domain was negative, reflecting frequent errors by the agent. However, as the agent continued to learn through interactions with the environment, the cumulative rewards improved, indicating reduced errors. Over subsequent episodes, the cumulative rewards approached zero with minor fluctuations, demonstrating convergence towards an optimal solution. This reward curve is influenced by the reward function definitions and environment setup.
The optimized toolpath, which successfully covers all defined points or states in the DRL environment, was achieved after 10,900 episodes. The efficiency of finding an optimized toolpath could potentially be enhanced by more judicious selection of environment parameters and reward definitions.


To evaluate the effectiveness of our DRL-optimized toolpath, we compared its grain morphology with that of a traditional zigzag path, which serves as the Base Case. The Base Case, figure \ref{case1}(a) includes figures showing both the conventional zigzag toolpath and the corresponding microstructure. In figure \ref{case1}(e), the histogram contrasts the grain morphology of the two toolpaths. The histogram displays the number of grains corresponding to each aspect ratio (AR) of both the zigzag path (blue bars) and the DRL-based path (orange bars) for same Hatch Space of 0.15 mm. The average ARs were $2.366$ for the zigzag path and $2.350$ for the DRL-based path. While the improvement in average AR appears modest, it is because the thermal model used for training didn't consider the residual heat caused by the tool path, thus the average grain ARs from different tool paths are similar. But this work provides an effective framework for designing complex tool paths, which is much more difficult to design than continuous process parameters such as laser power or scan speed. 

The histogram reveals that although the number of grains for AR values below 2.75 is similar for both toolpaths, there is a noticeable difference at higher AR values. Specifically, the DRL-based path reduces the number of larger AR grains compared to the zigzag path, indicating a more favorable grain morphology achieved with the DRL-optimized toolpath.

\subsubsection{Case 2: Using Grain Volume as a Reward Function}

\begin{figure}[H]
    \centering
    \includegraphics[width=0.95\linewidth]{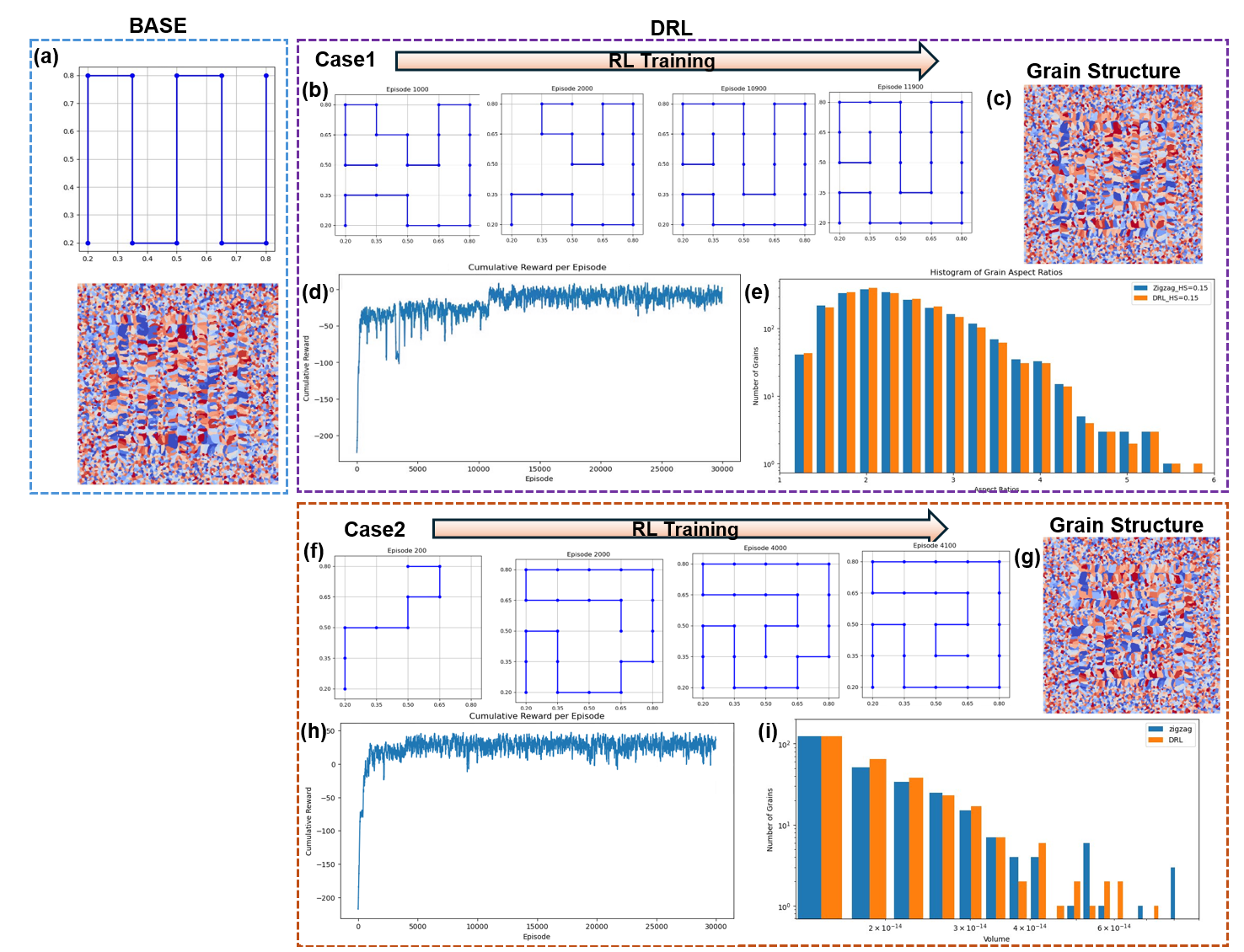}
    \caption{Comparison of DRL-optimized scan patterns using two reward functions: Case 1 (Average Aspect Ratio, AR) and Case 2 (Grain Volume, GV), with conventional zigzag scan as a baseline. Histogram shows modest gain in DRL against the base case.}
    \label{case1}
\end{figure}

\noindent In Figure \ref{case1}: Case 2, the training curve of the DRL method to minimize the grain volume (GV) is shown. This curve follows a similar trend to the previous DRL training curve. The generated toolpaths at different epochs are depicted in figure \ref{case1}(f). In this case, the agent was able to find the optimized path, visiting all the points in the grid, in just $4100$ episodes. This improvement is attributed to the different formulation of the positive reward function for average grain volume, as defined in Eq. \ref{eq:case_rewards}, compared to the average aspect ratio. This adjustment facilitated faster convergence, resulting in a higher cumulative reward than in the previous training.
The optimized toolpath and the corresponding microstructure for reducing GV are shown in figures \ref{case1}(f) and \ref{case1}(g). A histogram, figure \ref{case1}(i) comparing the grain morphology between the Zigzag path and the DRL-based path is also provided. The analysis indicates a significant improvement with the DRL-based toolpath, with a higher number of grains having a volume of \(2.5 \times 10^{-14}\) m\(^3\) and an increased number of grains with volumes greater than \(2.5 \times 10^{-14}\) m\(^3\). The average grain volumes in the melted region were \(8.8637 \times 10^{-15}\) m\(^3\) for the Zigzag path and \(8.8172 \times 10^{-15}\) m\(^3\) for the DRL-based path, demonstrating a notable improvement over the previous case.
For both cases, only the melted region was considered in the calculation of grain AR and GV, excluding the substrate regions. This method can also be adapted to optimize for higher AR and GV by modifying the corresponding reward functions.

\subsubsection{Case 3: Combining Aspect Ratio and Grain Volume}

\begin{figure}[H]
    \centering
    \includegraphics[width=0.95\linewidth]{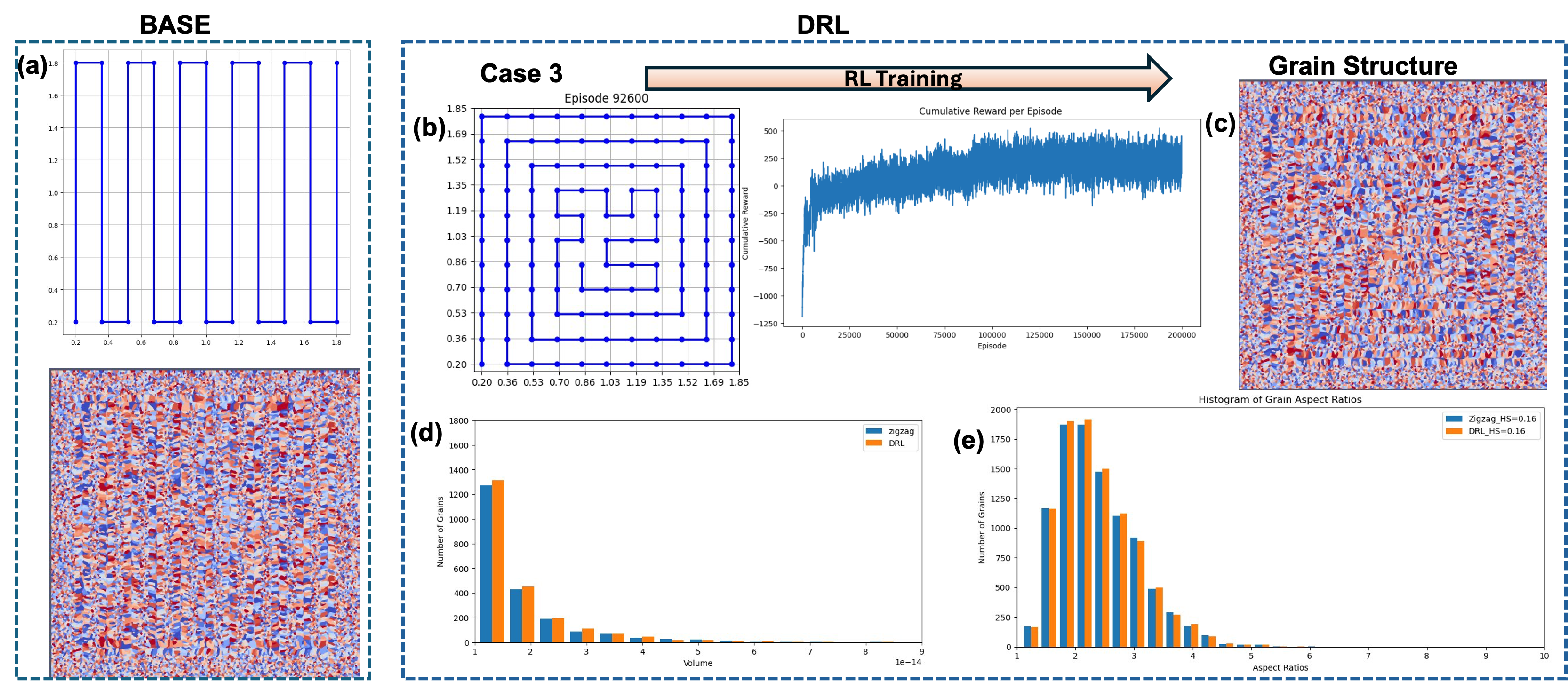}
    \caption{Case 3: DRL-optimized toolpath using combined aspect ratio and grain volume in the reward function with conventional zigzag scan as a baseline. DRL-generated path shows enhanced grain morphology by balancing AR and GV using the weighting parameters. }
    \label{case 3}
\end{figure}

\noindent Figure \ref{case 3} presents the results from Case 3, where a combined reward function was employed to simultaneously minimize both aspect ratio (AR) and grain volume (GV) for a larger domain of \(2 \text{ mm} \times 2 \text{ mm}\). The figures \ref{case 3}(a) is the conventional zigzag which is the base method, \ref{case 3}(b) is the toolpath generated from the DRL method and the corresponding cumulative reward and \ref{case 3}(c) is the generated microstructure. The performance of this approach was compared with the conventional Zigzag path, using the same hatch space of 0.16 mm. The results demonstrate a clear improvement over the smaller domain, with more distinct differences observed between the toolpaths.
From figures \ref{case 3}(d) and \ref{case 3}(e), it is evident that the DRL method yields a higher number of grains in regions of lower AR and GV compared to the traditional Zigzag toolpath. This combined approach proves more effective when optimizing toolpaths to reduce both AR and GV simultaneously. In this case, a weight of \( \alpha = 0.5 \) was used to prioritize lower AR, and a weight of \( \beta = 0.9 \) was assigned to emphasize lower GV. Although these weights have not been exhaustively explored, they offer a mechanism to control the desired properties through the combined reward function in the DRL method.

\section{METHODOLOGY}
\subsection{Direct Numerical Simulation: Phase-Field Method}

The phase field method is computational method for simulating grain evolution in metal solidification. A similar procedure to \cite{yang2021phase} has been followed in this study for the implementation. Order parameters $\eta_{i}(r,t)$ characterize grains in different orientation $i$, having a value of $\eta=1$ for grains at the $i^{th}$ orientation and transitioning to $\eta = 0$ for other grains. In this study , $i=20$ and grain boundary width $l = 9.6  \mu m$ chosen similar to \cite{moelans2008quantitative}. Further details has been discussed in supplementary material.
The governing equation that describes the evolution of the microstructure is represented by the time-dependent Ginzburg-Landau equation:

\begin{equation}
    \frac{\partial \eta_{i}(r,t)}{\partial t} = -L_g \frac{\delta F(\zeta, \eta_{i}, T)}{\delta \eta_{i}(r,t)}
\end{equation}
where $L_g$ signifies the kinetic coefficient associated with grain boundary mobility.
The total free energy $F$ is given by as:
\begin{equation}
    F = \int_V \left(m_g \left[ \sum_{i=1}^{n} \eta_{i} \left( \frac{\eta_{i}^4}{4} - \frac{\eta_{i}^2}{2} \right) + \gamma \sum_{i=1}^{n} \sum_{j\ne1}^{n} (\eta_{i})^{2} (\eta_{j})^{2}  + \frac{1}{4} + (1-\zeta)^{2} \sum_{i=1}^{n} (\eta_{i})^{2}\right] + \frac{\kappa_g}{2} (\nabla \eta_{i})^2\right) dV
\end{equation}

\subsubsection{Data Generation and Processing}
We performed six single-track, single-layer phase-field (PF) simulations using the process parameters outlined in \cite{choi4502223accelerating} to generate training data for a machine learning (ML) model. Neper 4.5.0  was utilized to create five distinct grain structures, each containing 30,000 grains within a $1 mm \times 0.3 mm \times 0.1 mm$ domain. The default Voronoi tessellation option resulted in  a mesh of 464 × 139 × 46 cells, with laser scanning ranging from $(0.15, 0.15, 0.1) mm$ to $(0.85, 0.15, 0.1) mm $. We varied the absorbed laser power from 20 W to 30 W in 2 W increments, while maintaining a constant scanning speed of 0.5 m/s. This setup ensured stable melt pool conditions without keyhole formation.

For the direct numerical simulation (DNS) in the first block of Figure \ref{fig-scheme}, we sampled a local Volume of Interest (VOI) measuring 0.17 mm × 0.17 mm × 0.07 mm at every 100-time step, following the laser path to capture the melt pool and grain evolution. This VOI size, augmented 19 times for rotation and permutation, was chosen for its efficiency in capturing the entire melt pool and for compatibility with the U-Net model's pooling steps. Detailed information on this data preparation process can be found in Figure \ref{fig-scheme}.

\subsection{The Reduced Order Model: 3D-UNET}
The aim of the machine learning is to predict the grain structure at the next time step based on current information about the grain structure and temperature data. A 3D image of the grain structure and temperature is used, with each voxel containing specific values. The semantic segmentation model U-Net is employed to make predictions, and a small region called the Volume of Interest (VOI) is selected to represent the grain structure at the current laser position. The VOI is concatenated with the temperature field at both the current and next time steps to form a 3-channel input image. To update the grain structure, only the microstructure within the VOI is employed, and the resulting microstructure is used in the next time step at a new location where a new VOI is extracted. This sequential approach allows the model to continuously progress in both space and time, enabling the simulation of the laser's impact on the grain structure during the process.

The 3D U-Net model, derived from the original 2D U-Net architecture \cite{ronneberger2015u}, was implemented using the PyTorch library. The model's input comprises a 3-channel image consisting of the initial grain orientations, initial temperature field, and the temperature field at 100 DNS time steps ahead. Importantly, each ML time step corresponds to 100 DNS time steps, allowing the ML model to operate without the critical time step requirement imposed by the explicit time integration scheme used in DNS. This coarse graining of the time step significantly enhances the efficiency of the ML model compared to the PF method.

In the encoder section of the U-Net, multiple double convolutions are performed before a max pooling operation. Each of these 3D convolutional layers transforms the input $\mathbf{x}$ by:
\begin{equation}
\mathbf{y}_{ijk} = \text{ReLU} \left( \sum_{m=-1}^{1} \sum_{n=-1}^{1} \sum_{o=-1}^{1} \mathbf{W}_{mno} \cdot \mathbf{x}_{i-o, j-n, k-o} + \mathbf{b} \right)
\end{equation}
where $\mathbf{W}$ and $\mathbf{b}$ are the convolutional kernel weights and bias, respectively. 
In this model, the Rectified Linear Unit (ReLU) function serves as the activation function, a common choice in deep neural networks. Each convolutional layer uses a kernel size of 3x3x3, a stride size of 1, and padding of 1, the latter being necessary to maintain the output image size, unlike the original U-Net architecture which lacks padding.
Batch normalization is applied after each convolutional layer to enhance training stability. Max pooling operations downsample the image by half using a pool size of 2x2x2 and a stride size of 2. In the decoder section of the U-Net, features are upsampled using the transposed convolution operator (or "up-convolution") and then concatenated with features from the encoder. These skip connections mitigate the vanishing gradient problem, enabling the retention of spatial information lost during downsampling. The second block ("blue") section of the overall pipeline in figure \ref{fig-scheme} gives details about the 3D-UNet architecture.

The final step in the process involves applying the log softmax function
\begin{equation}
S(y)_i = \frac{\exp(y_i)}{\sum_j \exp(y_j)}
\end{equation}

which is followed by the calculation of the cross-entropy loss

\begin{equation}
L(\hat{y}, y) = - \sum_j y_j \log(\hat{y}_j)
\end{equation}
The loss is then utilized to update the weights and biases during backpropagation. The output of U-Net is a 20-channel map, where each channel contains one of the 20 grain orientation order parameters.

\subsection{Deep Reinforcement Learning for Toolpath Design}
\begin{figure}[H]
    \centering
    \includegraphics[width=0.95\linewidth]{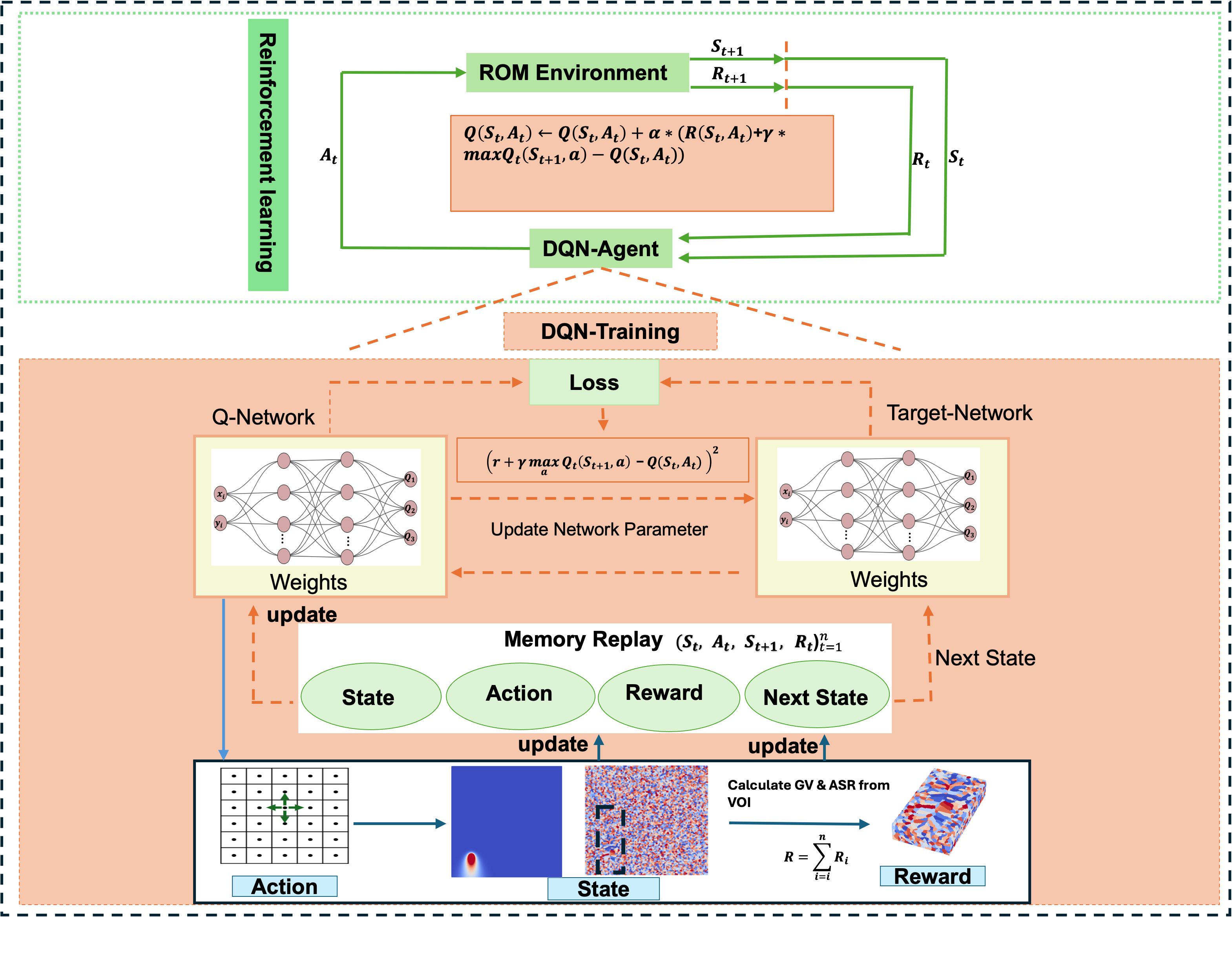}
    \caption{Deep Reinforcement Learning Framework Using a DQN Integrated with a 3D-UNet-based Reduced Order Model (ROM). The 3D-UNet generates ASR and GV metrics that serve as reward functions for the DQN, enabling the agent to learn an optimal control policy within the ROM environment.}
    \label{DRL-frame}
\end{figure}

In this section, we provide  details our proposed a Deep DRL based method that focuses on obtaining an optimized path minimizes the grain aspect ratio (AR) and grain volume (GV) which is desirable in manufactuing and other material properties parameters such as hatch space, laser power and velocity were kept constant. Our approach involves some basic reinforcement learning (RL) pipeline that includes defining environment, setting agent movement constraints, reward calculation and training. We used an extension of Q-Learning with deep neural network known as Deep Q-Network (DQN) \cite{drl1}. The general framework is depicted in figure \ref{DRL-frame} and it includes the following steps:\\

\noindent \textbf{1. Environment Definition}: The manufacturing domain was filled with uniformly distributed grid points. This grid environment serves as the space in which DQN agent operates. The grid was defined by the hatch space and actions available to the agent. To obtain the complete environment where the laser agent interacts, we coupled this with our reduce order model using the 3D-UNet define in section 3.2 which predicts the grain growth based on the laser coordinates (actions) at an accelerated rate. 

The DRL training for both  \textbf{Case 1} and \textbf{Case 2} , the manufacturing domain was filled 5x5 grid points with a hatch space of 0.15 and \textbf{Case 3}, the domain was filled with a 11x11 grid with a hatch space of 0.16. For all the cases, the laser agent always starts from the origin of a cartesian coordiante system. The agent can take one of four possible actions: \textbf{"Up" "Down", "Left" or "Right"} movements.  For the smaller domain (5x5 grid),  the state is represented by the position of the agent in the grid and this is encoded as a one-hot vector of length 25 by flattening the  5x5 grid where a 1 indicates the current position of the agent, and 0s indicate all other positions. For the larger domain (11x11 grid),  the state is represented by the position of the agent in the grid and this is encoded as a one-hot vector of length 121 by flattening the  11x11 grid where a 1 indicates the current position of the agent, and 0s indicate all other positions.
\newline
\newline
\textbf{2. Reward Function Definition}: Since the goal of the reward function is to minimize the aspect ratio and the grain volume of the predicted microstructure and avoid remelting which is influenced by the scan pattern, the reward function was designed in a way that it always encourages the agent to visit new points in the grid and avoid revisiting points or going out of bounds. 
If the agent moves to a new, unvisited point, it receives a positive reward using Eqn. \ref{eq:case_rewards}.
The environment consists of 25 equidistant fixed points for the smaller domain and 121 equidistant fixed points for the larger domain. The laser agent can move in four directions, resulting in a total of 100 possible movements for the smaller domain and 484 possible movements for the larger domain.

To avoid repetitive calculations and expedite the DRL training, the aspect ratios and grain volumes for each movement were pre-calculated using a previously trained 3D-UNet model. A real-time simulated aspect ratios and grain volumes were not used, as real-time simulations were too time-consuming due to the agent making identical movements multiple times over the episodes.
 If the agent moves out of bounds or revisits a point, it receives a penalty as defined by Eqns \ref{eq:r2} and \ref{eq:r3} and transitions to a terminal state, which signifies the end of the current episode and initiates a new episode. The agent receives a positive reward solely upon visiting a new point, with the reward quantified based on the average grain volume and aspect ratio associated with that movement. 
 In our study, three distinct reward cases were employed: the first case rewarded the agent for optimizing the toolpath with a focus solely on the aspect ratio; the second case provided rewards for reducing grain volume; and the third case combined both objectives by applying weighted factors ($\alpha$, $\beta$) to balance the optimization of aspect ratio and grain volume.
 
 
At the end of each episode, the agent incurs an additional penalty proportional to the number of unvisited points. The reward function for each episode is defined by Eqs. (11) to (15). Subsequently, the accumulated reward value, $R_{t}$, is calculated by summing all rewards obtained during that episode

\begin{equation}
    \text{Cumulative reward, } R = \sum_{i=1}^{4} R_i  \label{eq:cumulative_reward}
\end{equation}

\begin{equation}
    \begin{aligned}
        \text{Case 1: } \quad R_1 &= \frac{1}{\text{average aspect ratio}} \\
        \text{Case 2: } \quad R_1 &= \frac{1}{\text{average grain volume}} \\
        \text{Case 3: } \quad R_1 &= \alpha \left( \frac{1}{\text{average aspect ratio}} \right) + \beta \left( \frac{1}{\text{average grain volume}} \right)
    \end{aligned}
     \label{eq:case_rewards}
\end{equation}

\begin{equation}
    R_2 = -1  \label{eq:r2}
\end{equation}

\begin{equation}
     R_3 = -1 \label{eq:r3}
\end{equation}

\begin{equation}
    R_4 = -10 \times N_{p}  \label{eq:r4}
\end{equation}


where \( R_t \) denotes the accumulated reward, \( R_1 \) represents the reward for visiting a new point, calculated as the inverse of either the aspect ratio or the grain volume. \( R_2 \) and \( R_3 \) are constant penalties set to \(-1\), here, \( R_2 \) penalizes collisions with obstacles to discourage the agent from making moves that lead to collisions, and \( R_3 \) penalizes out-of-bounds movements to ensure the agent remains within the simulation domain. \( R_4 \) is a penalization parameter applied when the agent neglects certain areas of the computational domain. Here, \( N_p \) denotes the number of unvisited points. \\

\textbf{3. Deep Q-Network}: 
 The DQN algorithm is employed to learn the optimal policy for navigating the grid. Both the Q-network and the target network share the same architecture, consisting of two fully connected layers with $64$ units for the smaller domain (1mm by 1mm) and 128 units for the larger domain (2mm by 2mm). The Rectified Linear Unit (ReLU) activation function is applied to the hidden layers. The input layer has $25$ units for the smaller domain and $121$ units for the larger domain, representing the flattened state vector, while the output layer has $4$ units corresponding to the four possible actions.
A replay buffer is used to store the agent's experiences, including state, action, reward, and next state. We randomly sample mini-batches from this buffer for training the Q-networks,  mitigating the correlation between consecutive experiences, which results in more stable and efficient learning.
The target network parameters are updated periodically to stabilize training and are used to calculate the Q-values during training.

The Q-value function \( Q(s, a) \) estimates the expected cumulative reward obtained from taking action \( a \) in state \( s \) and following the optimal policy thereafter. The Q-value is updated iteratively using the Bellman equation \ref{eq:bell}. In this equation, \( r \) represents the reward received after taking action \( a \) in state \( s \), \( \gamma \) is the discount factor that determines the importance of future rewards, \( s' \) is the next state after taking action \( a \) in state \( s \), and \( a' \) is the action that maximizes \( Q(s', a') \) in the next state \( s' \). The update rule for \( Q(s, a) \) is given by:

\begin{equation}
Q(s, a) \leftarrow Q(s, a) + \alpha \left[ r + \gamma \max_{a'} Q(s', a') - Q(s, a) \right],
\label{eq:bell}
\end{equation}
where \( \alpha \) is the learning rate.

\noindent An epsilon-greedy strategy is used for action selection, where the agent's movement is directed by the maximum Q-values from the output layer. The Q-network is trained using the loss function defined in Eq \ref{eq:loss}  which measures the mean squared error (MSE) between the predicted Q-values and the target Q-values, as specified in Eq. \ref{eq:target}.

\begin{equation}
    \mathcal{L}(\theta) = \mathbb{E} \left[ \left( Q_{\text{target}} -Q_{\text{predicted}} \right)^2\right]
    \label{eq:loss}
\end{equation}
\begin{equation}
     Q_{\text{target}} = r + \gamma max_{a'} Q_{\text{target\_network}}(s', a')
     \label{eq:target}
\end{equation}

\noindent The agent explores the environment with probability \( \epsilon \) and exploits the learned policy with probability \( 1 - \epsilon \). The policy is shown in following equation:
\begin{equation}
a =
\begin{cases} 
\text{random action} & \text{with probability } \epsilon \\
\arg\max_{a'} Q(s, a') & \text{with probability } 1 - \epsilon 
\label{eq:explore}
\end{cases}
\end{equation}
To gradually reduce the exploration rate over time, epsilon decay is used. This approach ensures that the agent begins with a higher exploration rate to adequately learn about the environment, and progressively transitions towards exploitation as it gains confidence in its learned policy. The epsilon decay mechanism can be formalized as follows:
\begin{equation}
    \epsilon = \max(\epsilon \times \epsilon_{\text{decay}}, \epsilon_{\text{min}})
    \label{epsilon}
\end{equation}

\noindent This policy balances exploration, by selecting random actions, and exploitation, by choosing the best-known action. Initially, epsilon is set high to promote exploration and gradually decays over time to decrease exploration as the agent's understanding of the environment improves. The framework of the DRL method is illustrated in figure \ref{DRL-frame}.

\noindent This DRL algorithm seeks to learn the value of state-action pairs (Q-values) in order to derive an optimal policy for an agent interacting with the environment. The goal is for the agent to learn a policy that maximizes the cumulative reward over time.

\captionsetup{justification=centering,singlelinecheck=false}
\begin{table}[htbp]
\centering
\caption{Parameters Used in DRL Model Training} 
\label{process-parameters}
\begin{tabularx}{\linewidth}{XX}
\toprule
\textbf{Parameter} & \textbf{Value} \\
\midrule
Optimizer & ADAM \\
Learning Rate & 0.001 \\
Replay Buffer Size & 10000 \\
Discount Factor (\(\gamma\)) & 0.99 \\
Exploration Rate (\(\epsilon\)) & 1.0 \\ 
Minimum Epsilon & 0.01 \\
Epsilon Decay & 0.995 \\
\bottomrule
\end{tabularx}
\label{parameters:DRL}
\end{table}


\noindent Table \ref{parameters:DRL} summarizes the parameters and their respective values used for training the DRL model. The implementation was conducted using Python and the PyTorch library. Toolpaths were saved at every 100 episodes, and the cumulative reward curve was plotted to evaluate performance throughout the training process.

\section{Conclusion and Future Work}

In this work, we leveraged a 3D-UNet-based ROM as an efficient alternative to the computationally intensive PFM for simulating microstructure evolution in the L-PBF process. This ROM achieves a computational speedup of approximately two to three orders of magnitude, making it a viable substitute for exhaustive PFM simulations and enabling more efficient design of scanning strategies to achieve targeted microstructures. Our results demonstrate strong qualitative and quantitative agreement between the ROM and DNS, highlighting the model’s ability to accurately predict complex multi-track microstructures, even when trained solely on single-track PFM data. By integrating the ROM into a DRL framework, we successfully guided a laser control agent in optimizing scan patterns for enhanced microstructural outcomes.

The DRL framework, combined with the ROM, enables the generation of optimized scan paths at the mesoscopic scale, specifically minimizing average aspect ratios (AR) and grain volumes (GV) of the resulting microstructure. Our results show that this approach consistently outperforms traditional zigzag scanning methods, offering a more effective strategy for optimizing the L-PBF process. Overall, our DRL-based approach provides a scalable solution for managing the high degrees of freedom in scan pattern design, allowing real-time adjustments to minimize AR and GV, key factors affecting the final properties of LPBF-manufactured parts. This work emphasizes the potential of machine learning for optimizing toolpaths in additive manufacturing.

Looking forward, several promising avenues for future research arise from this work. Currently, the temperature solution used in our model is based on the steady-state Rosenthal solution; future studies could refine this by incorporating more dynamic thermal models that better capture transient thermal effects and their impact on microstructure evolution. Another direction could involve refining the DRL reward function to better align with specific material properties and process conditions. Finally, expanding the framework to handle more diverse and complex scan strategies may provide additional insights and further optimize L-PBF processes.

\section*{Author contributions statement}
\noindent
\textbf{Augustine Twumasi:} Conceptualization, Methodology, Formal analysis, Investigation, Software, Validation, Visualization, Writing—original draft, Writing—review \& editing.

\noindent
\textbf{Prokash Chandra Roy:} Methodology, Software, Formal analysis, Writing—review \& editing.

\noindent
\textbf{Zixun Li:} Data curation, Visualization, Writing—review \& editing.

\noindent
\textbf{Soumya Shouvik Bhattacharjee:} Investigation, Methodology, Writing—review \& editing. 

\noindent
\textbf{Zhengtao Gan:} Supervision, Conceptualization, Methodology, Investigation, Resources, Funding acquisition, Writing—review \& editing.

\section*{Declaration of Statements}
The authors declare no competing interest

\section*{Data Availability}
Data will be made available upon request.


\section*{Acknowledgements}
We gratefully acknowledge Jin Choi for his insightful discussions and contributions to this work. This research was supported by the U.S. Department of Energy (DOE) under Grant No. 226160684A: ADVANCED MULTI-PHYSICS MACHINE LEARNING FOR SUBSURFACE ENERGY SYSTEMS ACROSS SCALES

\bibliography{sample}

\end{document}